\documentclass[twocolumn,english,superscriptaddress]{revtex4-2}

\usepackage{mathpazo}

\usepackage[T1]{fontenc}
\usepackage[latin9]{inputenc}
\usepackage{color}
\usepackage{babel}
\usepackage{graphicx}
\usepackage{esint}
\usepackage[unicode=true,pdfusetitle,
 bookmarks=false,
 breaklinks=true,pdfborder={0 0 0},pdfborderstyle={},backref=false,colorlinks=true,pdfpagemode=FullScreen]
 {hyperref}
\hypersetup{citecolor=blue,filecolor=black,linkcolor=red,urlcolor=blue}
 
\begin{document}

\title{Inducing localized solutions via interaction in a self-defocusing
\\system with localized pumping}

\author{Wesley B. Cardoso}
\affiliation{Instituto de Física, Universidade Federal de Goiás, 74.690-900, Goiânia,
Goiás, Brazil}

\author{Mateus C. P. dos Santos}
\affiliation{Instituto de Ciências Tecnológicas e Exatas, Universidade Federal
do Triângulo Mineiro, 38064-200, Uberaba, Minas Gerais, Brazil}

\begin{abstract}
We investigate the emergence of induced localized coupled modes in
passive cavities with both loss and gain. Our model is based on linearly
coupled Lugiato-Lefever equations, where a Gaussian pump beam is applied
to only one mode. Through numerical simulations, we demonstrate that
self-defocusing systems can support the formation of localized stationary
modes in the partner field. The characteristics of these induced modes
are determined by key parameters, including coupling strength, cavity
decay rate, detuning effects, and the pump beam\textquoteright s intensity
and width.
\end{abstract}

\maketitle

\section{Introduction}

Solitons are a significant class of solutions emerging from nonlinear
wave equations, characterized by a precise equilibrium between dispersive
and nonlinear effects. These self-sustained waveforms preserve their
shape as they propagate and exhibit distinct nonlinear interactions
during collisions \citep{PhysRevLett.15.240,RevModPhys.61.763,kivshar2003optical}.
The prevalent presence of nonlinear wave equations in physics allows
solitons to emerge across multiple fields, such as water waves \citep{Craig_PL2006,Kodama_JPMT2010,Ablowitz_PRE12},
nonlinear fiber optics \citep{kivshar2003optical,hasegawa1995solitons,Erkintalo_PRL2011},
plasmas \citep{Lonngren_PP83,BERGE1998259,Shukla_NJOF2003}, Bose-Einstein
condensates (BECs) \citep{Burger_PRL99,Strecker_N02,Khaykovich_S02},
and degenerate fermion mixtures \citep{adhikari_EPJD06}. Experimental
observation has been the focus of extensive research, leading to its
realization in various physical systems, including vapors of alkali
metals \citep{Bjork_PRL74} and nonlinear waveguides \citep{Fleischer_N03,LEDERER20081}.

In particular, there is a class of solitons that emerge from dissipative
systems \citep{Purwins_AP10,Grelu_N12}, commonly found in nonlinear
optics. Although these models typically lack integrability, they provide
a more realistic description of various physical configurations. In
dissipative media, optical profiles are sustained by external pumping
in the form of laser beams. For passive optical systems, such solitons
are described by the Lugiato-Lefever (LL) equation \citep{Lugiato_PRL87}.
This equation has been widely used to model Kerr frequency combs \citep{Chembo_PRA13,Coen_NO13,Loures_PRL15,Pfeifle_PRL15,Hansson_NANO16},
contributing to advancements in spectroscopy \citep{Kippenberg_SC11},
optical metrology \citep{Cundiff_RSI01}, radio-frequency photonics
\citep{Torres_LPR14}, and optical communications \citep{Temprana_SC15}.

A natural extension of passive optical systems (spatial or temporal
domain) with gain and loss is to consider an optical system where
multiple electromagnetic modes interact with each other within a cavity
in a dissipative nonlinear medium. With pumping applied to each mode,
the localization and dynamics of the coupled solitons are described
by the set of coupled LL equations \citep{MARTIN201337,d2017coupled,Milian:18,PANAJOTOV2022112532}.
In these configurations, the modes are self-sustained through the
interplay of dispersion, diffraction, nonlinear effects, pumping,
dissipation, and their mutual interactions. In this context, the coupling
of electromagnetic modes in optical systems allows the study of vortex
solitons \citep{CAO2023112895}, vector cavity solitons \citep{Maitrayee_PRA20},
formation of dark dissipative solitons \citep{Kostet_PRA21}, and
the dynamics of high-order cavity solitons \citep{Sahoo_PRA19}. In
Ref. \citep{Fujii:17} the authors investigate the role of coupling
between clockwise and counterclockwise propagating modes in the generation
of Kerr frequency combs within whispering-gallery microcavities. Both
numerical and experimental analyses confirm that soliton formation
and comb power distribution are strongly influenced by the coupling
strength. The study employs coupled LL equations with a constant external
pump applied to a single mode to demonstrate Kerr comb generation.

Recently, the inclusion of a localized pump has shown great potential
for generating stable optical beams whose localization patterns directly
reflect the pump profile \citep{Cardoso_SR17,Cardoso_EPJD17,Kumar_S24,Santos_PRE25}.
Specifically, in Ref. \citep{Kumar_S24}, the authors introduce a
model of a passive optical cavity based on a two-dimensional variant
of the Lugiato-Lefever equations, incorporating a localized pump with
intrinsic vorticity $S$ and considering both cubic and cubic-quintic
nonlinearities. On the other hand, the induction of localized solutions
in partner fields -- including cases exhibiting symmetry breaking
-- solely due to coupling has been investigated in \citep{Hacker_S21,Santos_PRE21,Santos_ND23,Santos_ND25}.
However, previous studies on this phenomenon have been limited to
systems governed by nonlinear Schrödinger equations in the absence
of an external pump, leaving open the question of how a localized
pump may influence the coupling dynamics and the formation of localized
states in dissipative systems. In the present work, we examine the
impact of linear coupling on the localization of interacting optical
modes in a dissipative nonlinear cavity with spatially inhomogeneous
pumping applied to only one mode. Unlike previous studies, our focus
is on the induction of localized structures and the transfer of profile
characteristics mediated by the coupling mechanism. Additionally,
we investigate the influence of cavity decay rates and detuning, as
well as the modifications induced by variations in the pump beam\textquoteright s
spatial profile.

The remainder of this paper is structured as follows: the next section
introduces the theoretical model; Sec. \ref{sec:NUM} presents the
numerical results, detailing the characteristics and conditions for
the induction of localization; finally, Sec. \ref{sec:CONC} provides
the concluding remarks.

\section{Theoretical Model \label{sec:Theoretical-model}}

Consider a system that describes the dynamics of two interacting fields
in a localized pumping system, whose mathematical model is well described
by coupled LL equations, which are often used to model optical cavities
with dissipation and nonlinearity, given by:

\begin{equation}
\frac{\partial E_{1}}{\partial t}=\left[-(\alpha_{1}+i\Delta_{1})+i\frac{\partial^{2}}{\partial x^{2}}-i|E_{1}|^{2}\right]E_{1}+i\kappa E_{2}+F(x),\label{eq:LL1}
\end{equation}

\begin{equation}
\frac{\partial E_{2}}{\partial t}=\left[-(\alpha_{2}+i\Delta_{2})+i\frac{\partial^{2}}{\partial x^{2}}-i|E_{2}|^{2}\right]E_{2}+i\kappa E_{1},\label{eq:LL2}
\end{equation}
where $E_{1}(x,t)$ and $E_{2}(x,t)$ are complex fields, representing
the intensity profile of the two fields in time and space; $\alpha_{1}$
and $\alpha_{2}$ are damping rates of each field; $\Delta_{1}$ and
$\Delta_{2}$ are frequency shift (or phase adjustment) parameters
of fields 1 and 2, respectively; $\partial^{2}/\partial x^{2}$ is
the spatial diffusion term, representing the dispersion (or diffusion)
of the field along the spatial coordinate $x$; $|E_{1}|^{2}$ and
$|E_{2}|^{2}$ are the contributions due to Kerr-type nonlinearities;
$\kappa$ represents the interaction between the fields $E_{1}$ and
$E_{2}$, allowing them to mutually influence their dynamics. $F(x)$
is the external pumping applied exclusively to the field $E_{1}$.

Note that the nonlinearity of the system is self-defocusing for both
fields, which means that there can be no equilibrium between the dispersion/diffusion
and the nonlinear effects of the system (i.e., no localized solution)
in the absence of pumping. In this study, we will investigate the
influence of the pumping term on the localization of field 1 and how
the interaction between the fields induces localization in the partner
field.

As an example of localized pumping, in this work, we will employ a
Gaussian pump described by the following expression:
\begin{equation}
F(x)=\frac{P_{0}}{\sqrt{\pi}W}\exp\left(-x^{2}/W^{2}\right),\label{eq:pump}
\end{equation}
where $P_{0}$ and $W$ represent the pump intensity and width, respectively.

\section{Numerical results \label{sec:NUM}}

In this section, we present the numerical results derived from Eqs.
(\ref{eq:LL1}) and (\ref{eq:LL2}), with a localized pump applied
to field $E_{1}$ as specified in Eq. (\ref{eq:pump}). The numerical
integration was performed using a pseudo-spectral method, where the
diffusive term in $x$ was addressed using the fast Fourier transform,
and the temporal evolution was solved using the 4th-order Runge-Kutta
method. We employed $256$ spectral points over a spatial domain of
size $40$, with a time step of $10^{-3}$. The system was initialized
with both fields in the vacuum state and evolved until convergence
of the localized solution was achieved. For more details on this methodology,
we direct the reader to the reference book \citep{Yang_10}.

\begin{figure*}[tb]
\centering \includegraphics[width=0.9\textwidth]{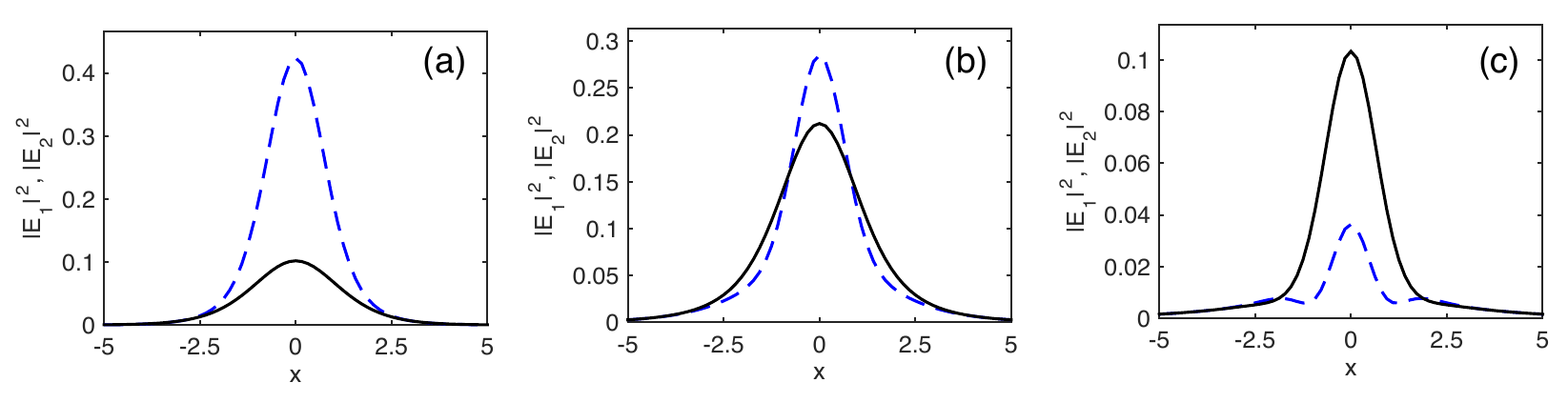}

\caption{Numerical profile of the intensities ($|E_{1}|^{2}$ and $|E_{2}|^{2}$,
represented by dashed and solid lines, respectively) of the localized
states after their respective stabilization for three different values
of the coupling parameter, specifically (a) $\kappa=1$, (b) $\kappa=2$,
and (c) $\kappa=5$. The system starts from the vacuum as the initial
state. The other parameter values are $\Delta_{1}=\Delta_{2}=\alpha_{1}=\alpha_{2}=W=1$
and $P_{0}=3$.}

\label{F1}
\end{figure*}

\begin{figure*}[tb]
\centering 
\includegraphics[width=0.3\textwidth]{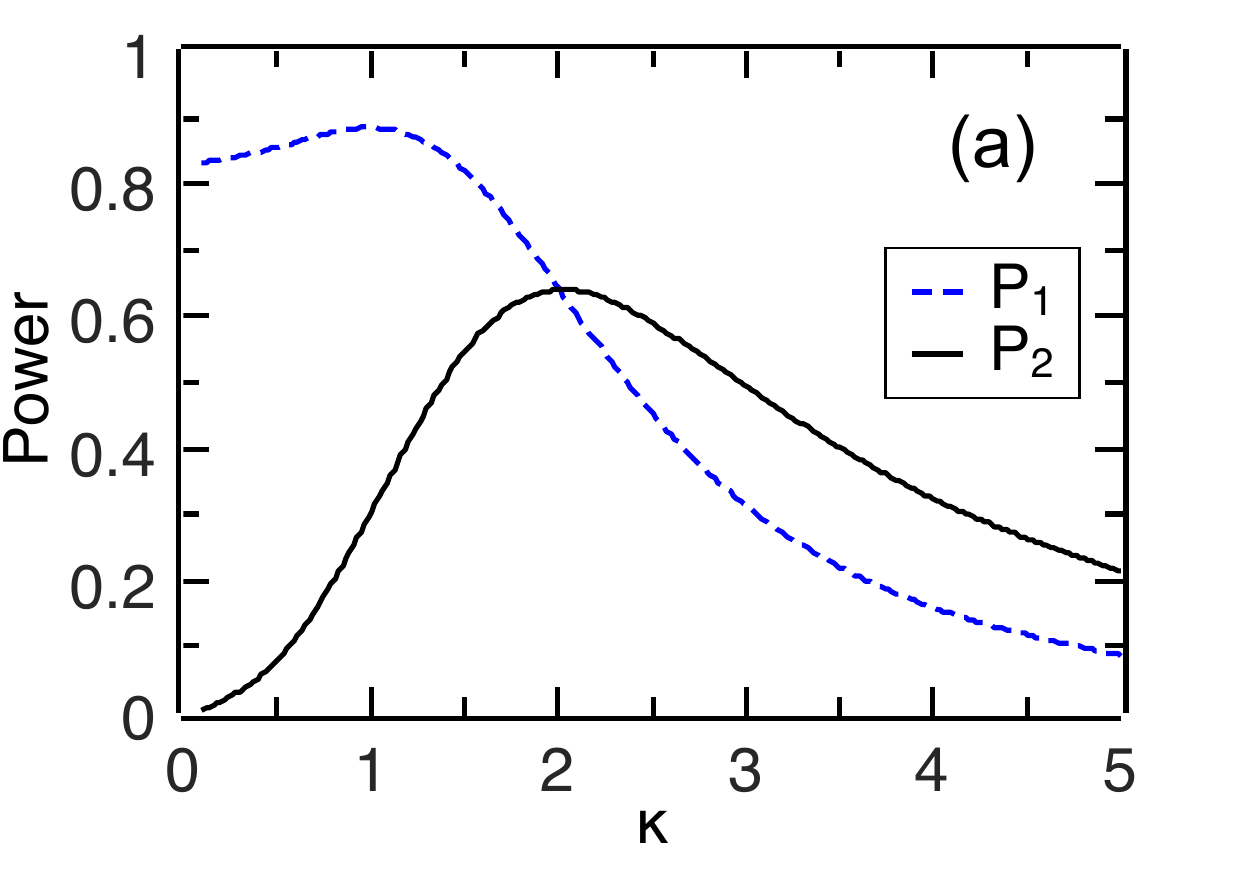} 
\includegraphics[width=0.3\textwidth]{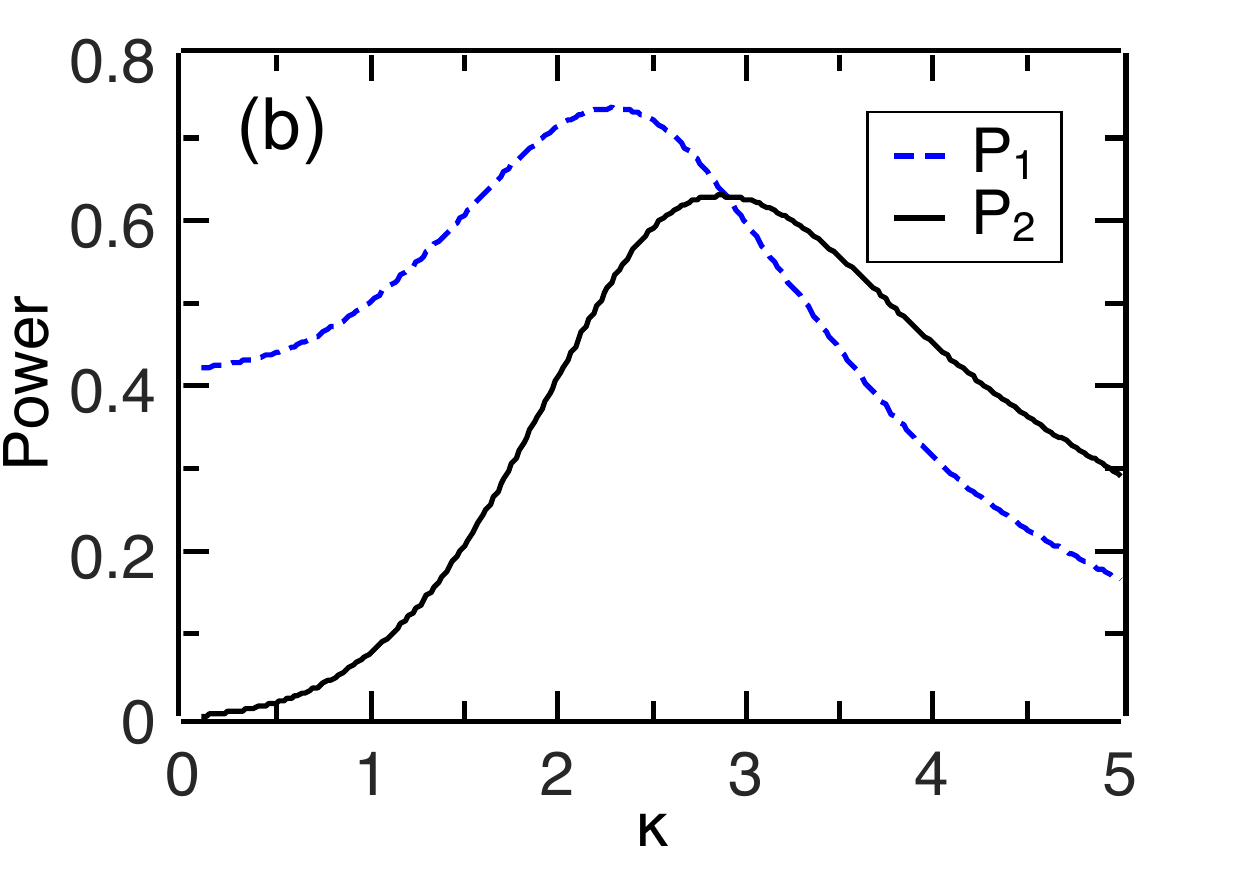}
\includegraphics[width=0.3\textwidth]{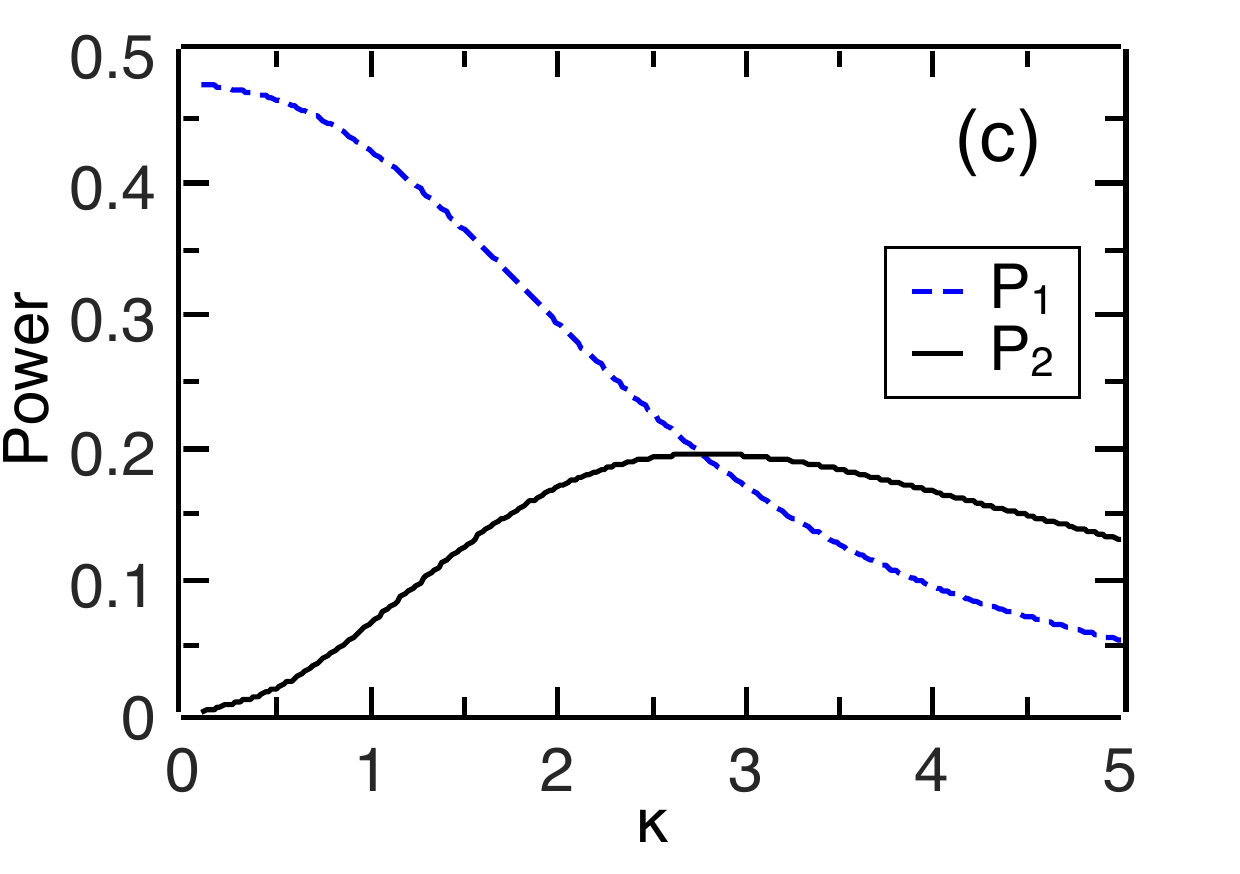}

\caption{Power of the fields $E_{1}$ and $E_{2}$ as a function of $\kappa$.
In panel (a), the parameter values are the same as those used in Fig.
\ref{F1}. In panel (b), we modify the values to $\Delta_{1}=\Delta_{2}=2$
(keeping the other parameter values the same). In panel (c), we set
$\alpha_{1}=\alpha_{2}=2$ and keeping the other parameter values
the same as in Fig. \ref{F1}.}

\label{F2}
\end{figure*}

\textbf{\emph{Influence of coupling parameter $\kappa$}}: First,
we will examine the influence of the coupling parameter on the localization
of the partner field. To this end, Fig. \ref{F1} presents an example
of the intensity profiles of the fields for three different values
of this parameter, while keeping all other system parameters fixed.
It can be observed that, as the coupling parameter increases, the
partner field exhibits greater power, eventually surpassing the power
of field $E_{1}$. Indeed, we find that for the same values of the
other parameters used in Fig. \ref{F1}, the powers of both fields
become equal when $\kappa$ is approximately $2.0$.

To further illustrate this, Fig. \ref{F2}(a) shows the relationship
between the powers of the fields and the coupling parameter $\kappa$.
It is worth noting that the power values begin to decrease after a
certain $\kappa$ threshold. Moreover, the maximum power for field
$E_{2}$ is reached when it equals the power of field $E_{1}$. In
another simulation, we observed that increasing the detuning shifts
the peak of the power curve to higher values of $\kappa$. This result
is illustrated in Fig. \ref{F2}(b), where we set $\Delta_{1}=\Delta_{2}=2$,
while keeping the other parameter values the same as those used to
produce Fig. \ref{F2}(a). Note that the values of the two powers
now become equal for $\kappa\simeq2.9$. Next, in Fig. \ref{F2}(c),
a similar behavior can be observed, but now considering a higher decay
rate, i.e., $\alpha_{1}=\alpha_{2}=2$, while keeping the other parameter
values the same as those used in Fig. \ref{F1}. In this case, the
power values coincide when $\kappa\simeq2.8$.

\begin{figure*}[tb]
\centering 
\includegraphics[width=0.3\textwidth]{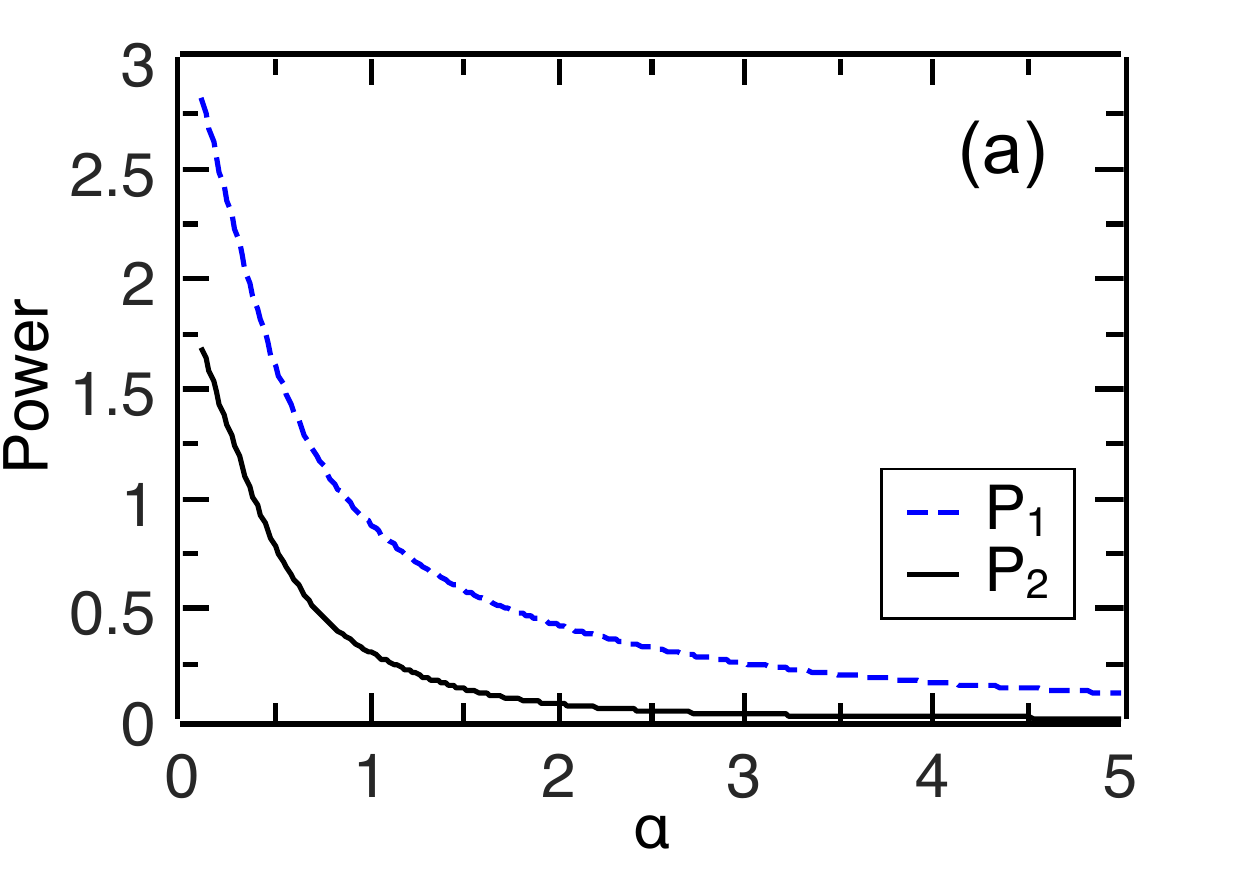} 
\includegraphics[width=0.3\textwidth]{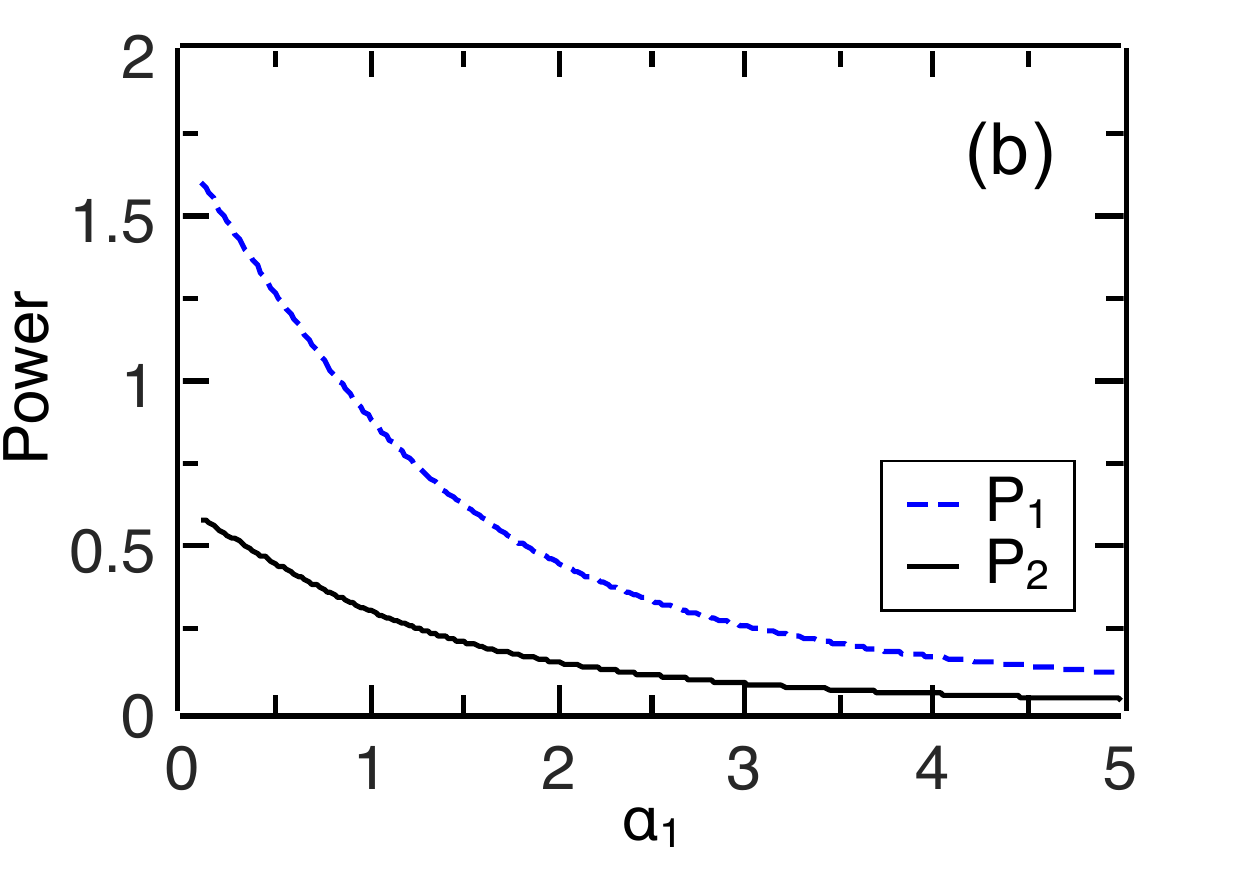}
\includegraphics[width=0.3\textwidth]{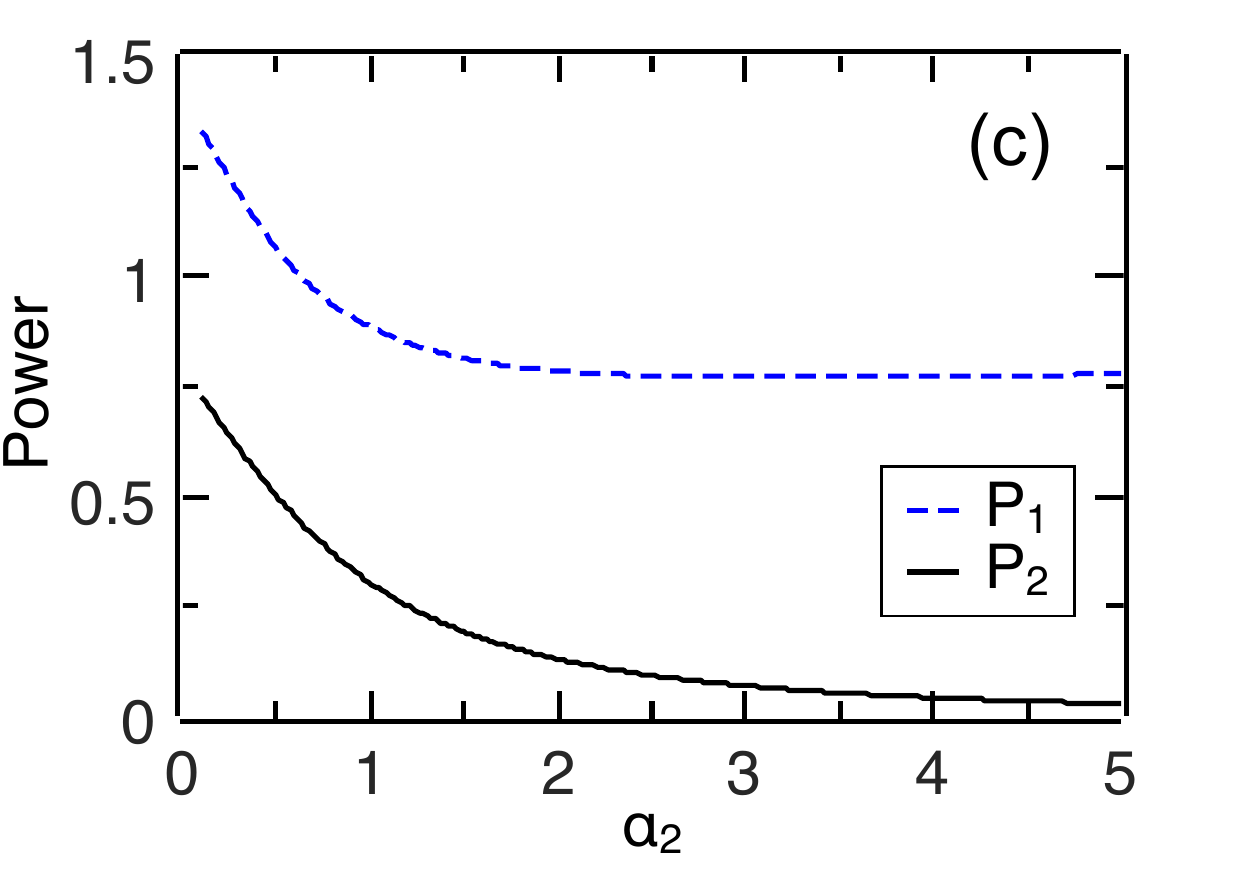}

\caption{Power of the fields $E_{1}$ and $E_{2}$ as a function of decay rate.
(a) Symmetric case, considering $\alpha_{1}=\alpha_{2}=\alpha$; (b)
Field $E_{2}$ with a fixed decay rate $\alpha_{2}=1$; (c) Field
$E_{1}$ with a fixed decay rate $\alpha_{2}=1$. The other parameter
values were $\Delta_{1}=\Delta_{2}=\kappa=W=1$ and $P_{0}=3$.}

\label{F3}
\end{figure*}

\textbf{\emph{Influence of decay rates}} \textbf{$\alpha_{1,2}$}:
In order to investigate the influence of the decay parameter on the
induction of localization in the partner field ($E_{2}$), we first
analyze the behavior of the power for the symmetric case of this parameter,
i.e., choosing equal values for the decay rates $\alpha_{1}=\alpha_{2}=\alpha$.
The result of our simulation is shown in Fig. \ref{F3}(a), where
it can be observed that the higher the value of $\alpha$, the lower
the power of both fields. Moreover, the power of the induced field
rapidly decreases to values close to zero, reaching $P_{2}\simeq0.004$
when $\alpha=5$.

In the result shown in Fig. \ref{F3}(b), we kept the decay rate of
field $E_{2}$ fixed at $\alpha_{2}=1$, while varying the value of
$\alpha_{1}$. We observed a similar behavior to that presented in
Fig. \ref{F3}(a), indicating that the decay rate of field $E_{1}$
significantly impacts the response of the induced field $E_{2}$.
On the other hand, to examine the impact of the decay rate of field
$E_{1}$, in Fig. \ref{F3}(c) we kept it fixed at $\alpha_{1}=1$
while varying the value of $\alpha_{2}$. It can be seen that as $\alpha_{2}$
increases, the power of field $E_{2}$ approaches zero, while the
power of field $E_{1}$ stabilizes at a value close to $0.78$. Therefore,
when the ratio $\alpha_{1}/\alpha_{2}$ becomes very large, the induction
effect is lost, causing field $E_{2}$ to vanish while maintaining
the localized structure of field $E_{1}$.

\begin{figure*}[tb]
\centering 
\includegraphics[width=0.3\textwidth]{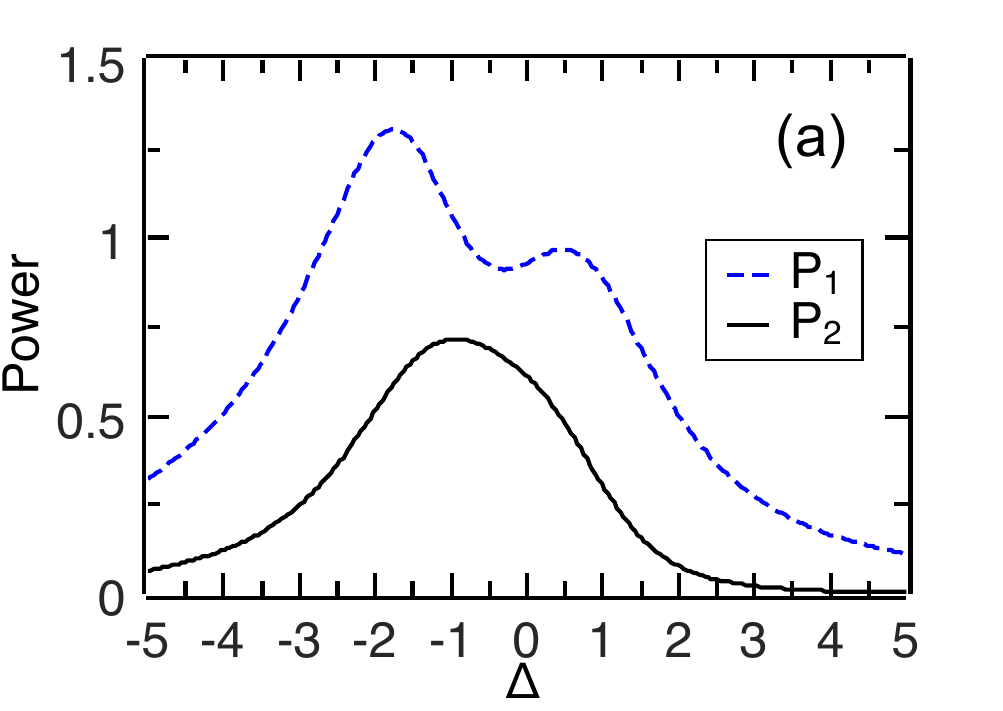} 
\includegraphics[width=0.3\textwidth]{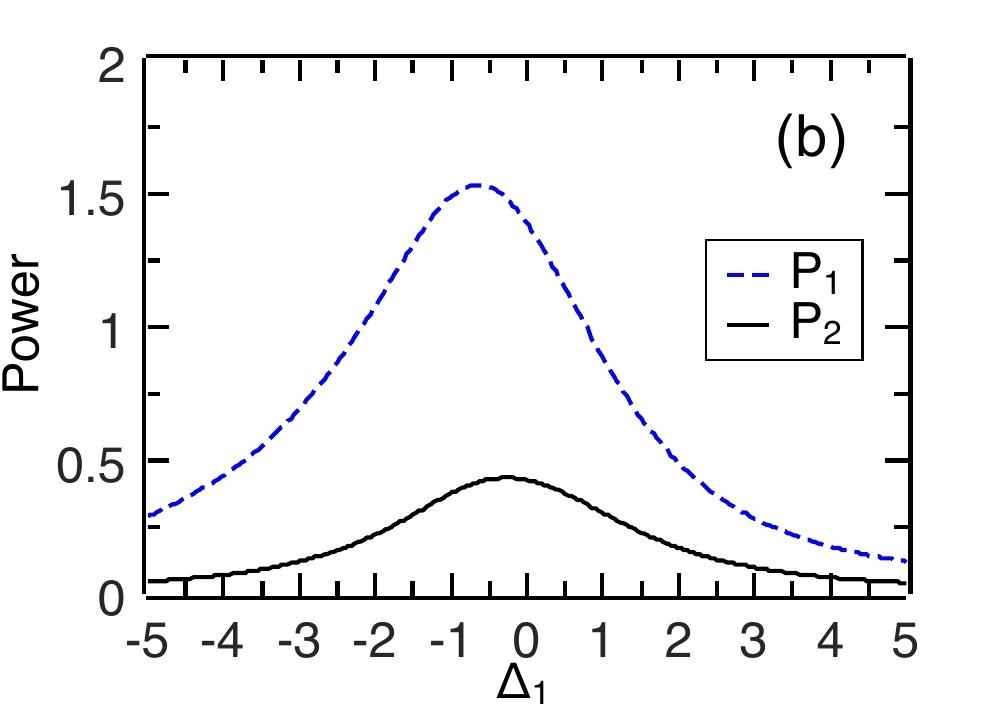}
\includegraphics[width=0.3\textwidth]{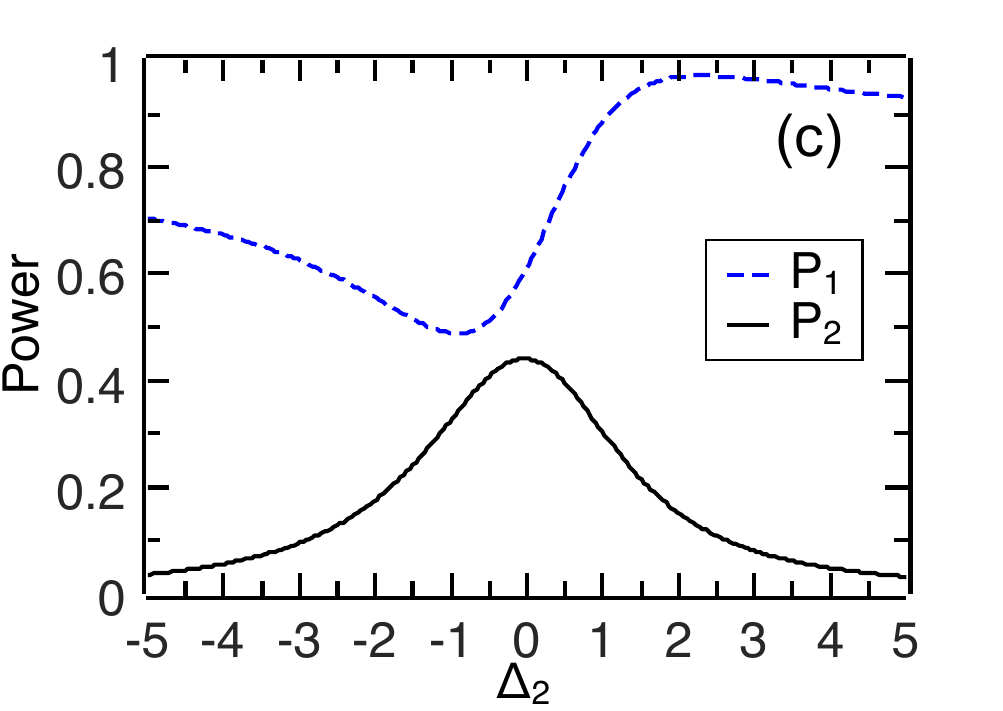}

\caption{Power of the fields $E_{1}$ and $E_{2}$ as a function of the frequency
shift. In panel (a), the symmetric case is presented, with $\Delta_{1}=\Delta_{2}=\Delta$.
The variation of $\Delta_{1}$ only is shown in panel (b), where we
fix $\Delta_{2}=1$. In panel (c), we reverse the setup, fixing $\Delta_{1}=1$
while varying $\Delta_{2}$. The other parameter values were $\alpha_{1}=\alpha_{2}=\kappa=W=1$
and $P_{0}=3$.}

\label{F4}
\end{figure*}

\textbf{\emph{Frequency Detuning Effect $\Delta$}}: In Fig. \ref{F4},
we present the analysis of the impact resulting from changes in the
detuning. Specifically, in Fig. \ref{F4}(a), we consider the symmetric
case, where both fields have the same detuning ($\Delta_{1}=\Delta_{2}=\Delta$).
We can observe that the power is higher when a redshift is applied
(negative values of $\Delta$). For field $E_{1}$, two prominent
power peaks occur at $\Delta\simeq-1.8$ and $\Delta\simeq0.5$, with
the former exhibiting a higher power. The induced field, on the other
hand, exhibits a single power peak at $\Delta=-0.9$.

In Fig. \ref{F4}(b), we analyze the influence of detuning solely
for field $E_{1}$, keeping $\Delta_{2}=1$ fixed. Interestingly,
unlike the symmetric case shown in Fig. \ref{F4}(a), the power of
field $E_{1}$ now exhibits a single peak at $\Delta_{1}\simeq-0.7$,
while the power peak of $E_{2}$ occurs at $\Delta_{1}\simeq-0.25$.
Finally, in Fig. \ref{F4}(c), we observe the behavior of the powers
as a function of $\Delta_{2}$ only, with $\Delta_{1}=1$. Here, we
can see that $\Delta_{2}$ introduces a significant change in the
power of the induced solution, reaching a maximum at $\Delta_{2}=-0.1$.
Additionally, the power of field $E_{1}$ also undergoes more pronounced
changes when $\Delta_{2}$ is close to zero.

\begin{figure}[tb]
\centering 
\includegraphics[width=0.48\columnwidth]{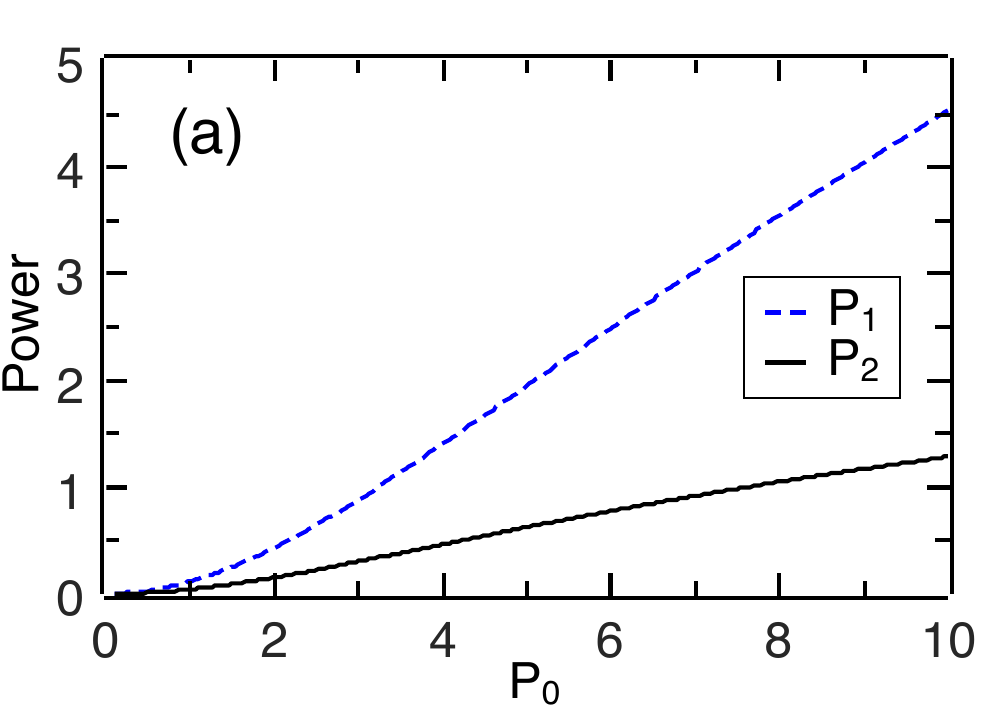} 
\includegraphics[width=0.48\columnwidth]{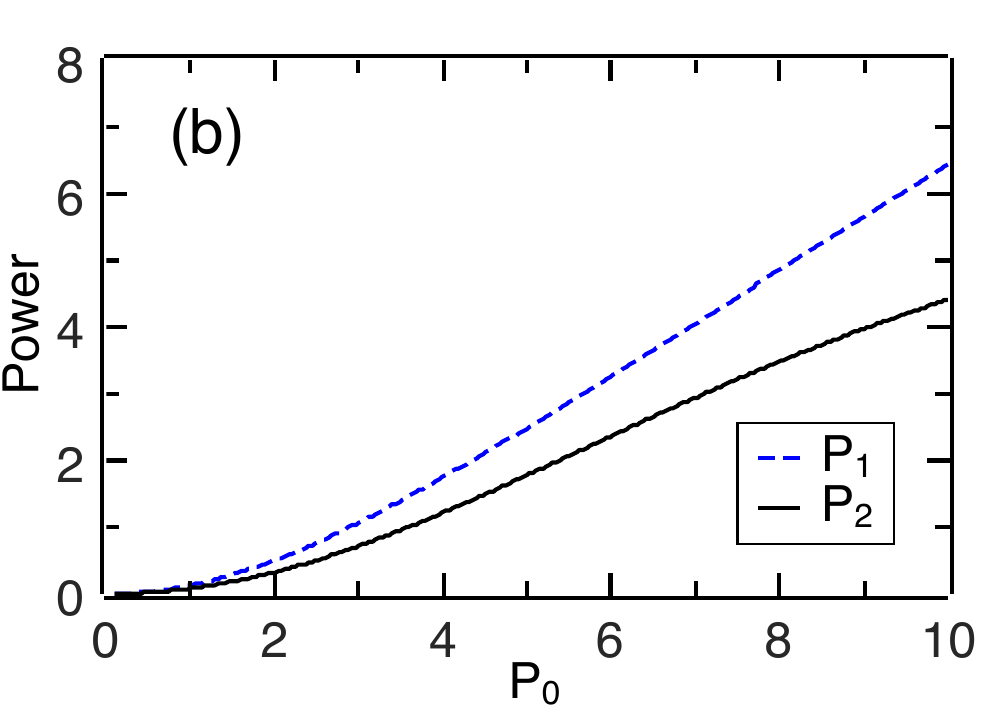}

\caption{Power as a function of the pump intensity $P_{0}$, considering (a)
$\Delta_{1}=\Delta_{2}=1$ and (b) $\Delta_{1}=\Delta_{2}=-1$. The
other parameter values used here were $\alpha_{1}=\alpha_{2}=\kappa=W=1$.}

\label{F5}
\end{figure}

\textbf{\emph{Influence of Pump Amplitude $P_{0}$}}: We now analyze
the changes in the induced solution when considering different values
of the pump amplitude. In this regard, Fig. \ref{F5} shows how the
power of solutions $E_{1}$ and $E_{2}$ varies with changes in this
parameter. Specifically, in Fig. \ref{F5}(a), we consider a positive
detuning (blueshift), while in Fig. \ref{F5}(b), we examine the opposite
case. Note that the behavior of the powers is similar in that both
increase as the pump amplitude rises. Moreover, when the detuning
is negative (Fig. \ref{F5}(b)), the induced field exhibits a larger
power increase compared to the case with positive detuning (Fig. \ref{F5}(a)),
as previously observed in Fig. \ref{F4}. However, the gap between
the powers in the negative detuning case is smaller, as can be visually
observed in Fig. \ref{F5}.

\begin{figure}[tb]
\centering 
\includegraphics[width=0.48\columnwidth]{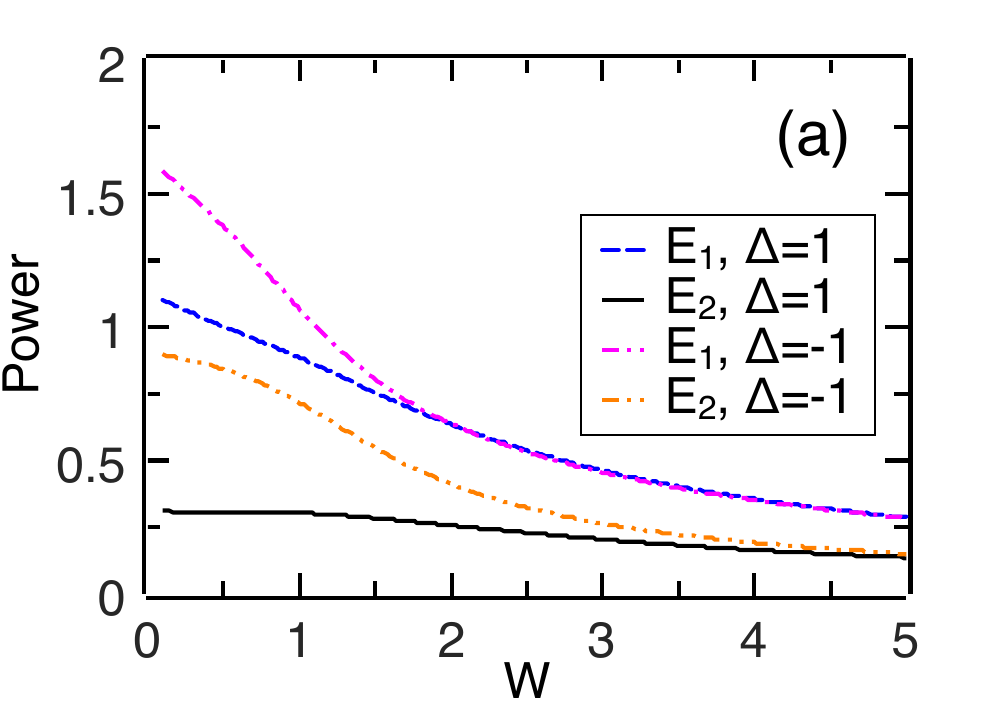} 
\includegraphics[width=0.48\columnwidth]{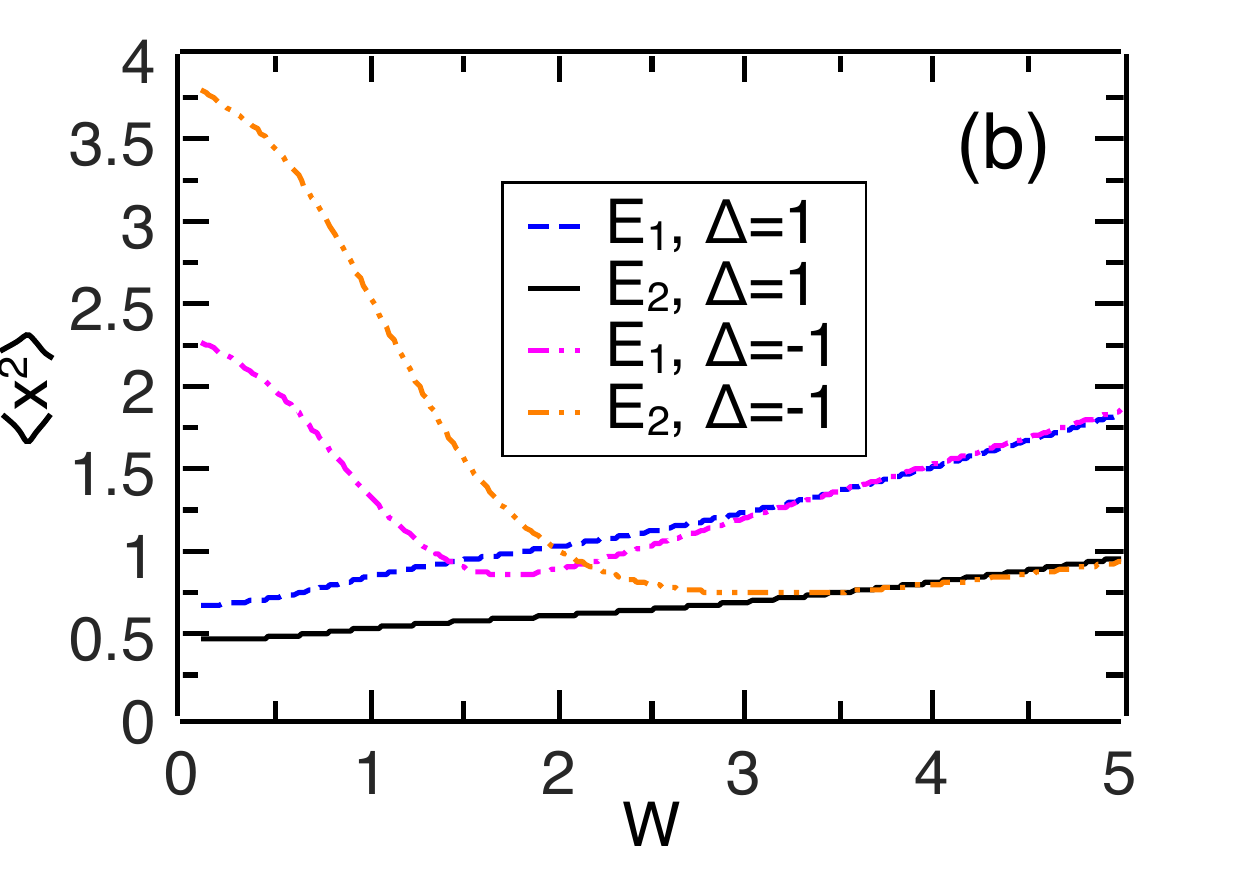}

\caption{(a) Power and (b) average width of the solutions for fields $E_{1}$
and $E_{2}$ as a function of the pump width $W$, for two distinct
values of detuning. The other parameter values used here were $\alpha_{1}=\alpha_{2}=\kappa=1$
and $P_{0}=3$.}

\label{F6}
\end{figure}

\textbf{\emph{Influence of Pump Width $W$}}: Our final study involves
analyzing the dependence of the pump width on the induction of localization
in the partner field $E_{2}$. A summary of the numerical results
is presented in Fig. \ref{F6}, where the power of the solutions $E_{1}$
and $E_{2}$ is shown in Fig. \ref{F6}(a), while the corresponding
average width is displayed in Fig. \ref{F6}(b). To calculate the
average width, we used the definition 
\[
\langle x^{2}\rangle=\int_{-\infty}^{\infty}x^{2}|E_{k}|^{2}dx,
\]
where the index $k=1$ or $2$ corresponds to the calculation of the
average width for field $E_{1}$ or $E_{2}$, respectively. Furthermore,
these figures illustrate the results of our simulations for two distinct
detuning values, namely $\Delta_{1}=\Delta_{2}=1$ and $\Delta_{1}=\Delta_{2}=-1$.

In Fig. \ref{F6}(a), it can be observed that the power of the solutions
decreases as the pump width increases. Additionally, it is noteworthy
that as $W$ increases, the difference between the power obtained
for both detuning values diminishes, becoming indistinguishable beyond
a certain value of $W$. This result is also observed when examining
the average width shown in Fig. \ref{F6}(b), where the values become
indistinguishable at approximately $W>3.5$, for both fields.

Another interesting result pertains to the relationship between power
and average width for small values of $W$. Note that the power of
field $E_{1}$ is always greater than that of field $E_{2}$. However,
for the negative detuning value, we observe that the width of field
$E_{2}$ is larger than that of $E_{1}$, which is the opposite of
what occurs when the detuning is positive. Furthermore, when the detuning
is negative, there is a point where the widths of both solutions are
equal, corresponding to $W\simeq2.15$.

\section{Conclusion \label{sec:CONC}}

In conclusion, we investigated the induction of localized solutions
in a self-defocusing nonlinear optical system with localized pumping,
emphasizing the role of linear coupling in shaping the formation of
structured optical fields. Through numerical simulations, we analyzed
the influence of key system parameters, including the coupling strength
$\kappa$, decay rates $\alpha_{1}$, $\alpha_{2}$, frequency detuning
$\Delta$, pump amplitude $P_{0}$, and pump width $W$. Our results
demonstrate that the interaction between the coupled fields $E_{1}$
and $E_{2}$ plays a fundamental role in defining the localization
dynamics, with the induced stationary modes exhibiting distinct characteristics
depending on the coupling strength and the spatial properties of the
pump. 

Notably, our findings reveal that, even in the absence of direct excitation,
localized states can emerge in the partner field solely due to the
coupling mechanism. This behavior is strongly influenced by the interplay
between detuning and dissipation, which governs the stability and
spatial distribution of the induced modes. 

Our study extends previous investigations on coupled nonlinear optical
systems by incorporating the effects of a spatially inhomogeneous
pump, highlighting new possibilities for tailoring localized states
through controlled interactions. These results contribute to the broader
understanding of dissipative solitons and structured light in nonlinear
cavities, with potential applications in photonic information processing
and optical trapping. 
\begin{acknowledgments}
We acknowledge the financial support provided by the Brazilian agencies
CNPq (grant \#306105/2022-5 and Sisphoton Laboratory- MCTI No. 440225/2021-3),
CAPES, and FAPEG. This work was also performed as part of the Brazilian
National Institute of Science and Technology (INCT) for Quantum Information
(Grant No. 465469/2014-0).
\end{acknowledgments}

\bibliographystyle{apsrev4-2}
\bibliography{REFS}

\end{document}